\documentclass{article}

\usepackage{INTERSPEECH2020}
\usepackage[table]{xcolor}
\usepackage{enumitem}
\usepackage{hyperref}
\title{Data augmentation using prosody and false starts to recognize \\non-native children's speech}
\name{Hemant Kathania, Mittul Singh, Tam\'as Gr\'osz, Mikko Kurimo}
\address{
 Department of  Signal Processing and Acoustics, Aalto University, Finland}
\email{firstname.lastname@aalto.fi}
\begin{document}
%
\maketitle
\begin{abstract}
This paper describes AaltoASR's speech recognition system for the INTERSPEECH 2020 shared task on Automatic Speech Recognition (ASR) for non-native children's speech. The task is to recognize non-native speech from children of various age groups given a limited amount of speech. Moreover, the speech being spontaneous has false starts transcribed as partial words, which in the test transcriptions leads to unseen partial words. To cope with these two challenges, we investigate a data augmentation-based approach. Firstly, we apply the prosody-based data augmentation to supplement the audio data. Secondly, we simulate false starts by introducing partial-word noise in the language modeling corpora creating new words. Acoustic models trained on prosody-based augmented data outperform the models using the baseline recipe or the SpecAugment-based augmentation. The partial-word noise also helps to improve the baseline language model. Our ASR system, a combination of these schemes, is placed third in the evaluation period and achieves the word error rate of 18.71\%. Post-evaluation period, we observe that increasing the amounts of prosody-based augmented data leads to better performance. Furthermore, removing low-confidence-score words from hypotheses can lead to further gains. These two improvements lower the ASR error rate to 17.99\%.
\end{abstract}
\section{Introduction}
\label{sec:intro}
Many schools around the world teach English as a foreign language to their pupils. As the trend grows, it creates a demand to develop an objective and reliable English language skills assessment for young learners. In turn, the assessment system requires a robust speech recognition system. However, recognizing children's speech is challenging and the non-native nature of the speech adds further complexity to this task.

INTERSPEECH 2020 shared task offers a unique opportunity to recognize non-native children's speech. As part of the shared task, the provided dataset includes various speech phenomena like code-switching between multiple languages, a large number of spontaneous speech phenomena (like hesitations, false starts, fragments of words), presence of non-collaborative speakers (students often joke, laugh, speak softly, etc.) and multiple age-groups (9 - 16 years). Handling these phenomena is exacerbated by the presence of background noise in the dataset. Furthermore, the dataset provides only limited amounts of labeled audio ($\sim50$ hours). 

In this paper, we aim to handle the data scarcity, acoustic variability and false starts in text for building an Automatic Speech Recognition (ASR) system. To alleviate data scarcity, we augment training audio data.  Specifically, we compare SpecAugment- \cite{ParkCZCZCL19} and prosody-based \cite{8049381} data augmentation (section \ref{sec:data_augment}). SpecAugment, recently popularized for building a robust ASR, has not been explored for processing children's speech. This technique ignores the prosodic variability of children's speech from different age-groups, which is relevant to this task. In contrast, prosody-based data augmentation leverages this acoustic variability to increase the amount of data. Our expriments also show that the prosody-based augmentation is more beneficial than SpecAugment for children's speech.

The shared task represents false starts in speech as partial words. Approximately 10\% of the development and evaluation set utterances contain such words. Prior work \cite{PrylipkoVSW12,7804538} has noted that handling disfluencies like false starts can lead to better prediction of the next word. To handle the partial words, we randomly add partial words by splitting existing words to the language modeling text (section \ref{ssec:partial_word}). We observe that noising text in this manner improves the ASR performance.

During the evaluation period, our ASR system ranked third, using a combination of prosody- and SpecAugment-based data augmentation while handling false starts (section \ref{sec:sys_comb}). We also release tools for prosody-based augmentation and handling false starts publicly \footnote{\url{https://github.com/kathania/Interspeech-2020-Non-native-children-ASR.git}}. Post-evaluation period, we experiment with increasing the augmented data, which improves the performance of prosody-based ASR, whereas the performance of SpecAugment-based ASR drops (section \ref{sec:data_augment}). We also implement a filtering scheme to remove words with low confidence scores (section \ref{sec:conf_filter}). The filtering technique can deal with non-English words, which are discounted by the shared task metric. The filtering scheme helps to improve the performance of the combined system further.
\renewcommand{\arraystretch}{1.6}
\begin{table}[!t]
\centering
\caption{\label{tab:corpus}The table reports statistics for the organizer provided datasets used in our paper: two training sets: train-1 and train-2, the development set (dev) and the evaluation set (eval).}
\vspace{-0.3cm}
\begin{tabular}{|l|c|c|c|c|c|c|}\hline
Corpus 			&{train-1}	&{train-2} & dev & eval	\\ \hline \hline 
No. of words 		& 22450	& 136578	 	& 5287	& 6206		\\ \hline
No. of pupils 		& 338	& 3112	 	& 84	& 84		\\ \hline
Duration (hours)	 	& 8.59		& 40.29		& 2.05		& 2.20		\\ \hline
\end{tabular}
\vspace{-0.6cm}
\end{table}

\section{ASR for non-native children's speech}
\subsection{Dataset}
\label{ssec:data}
In this challenge, the organizers provide us with an English portion of speech dataset called TLT-school corpus \cite{TLT_2020}. The corpus consists of audio from Italian pupils speaking English collected between the years 2016 and 2018. The pupil's ages range from 9 to 16 years, belonging to four different school grade levels. The pupils are divided into three age-based groups A1 (9-10), A2 (12-13) and B1 (14-16). Each age-group is asked to answer questions according to their language skills. The recorded answers form the speech provided in this challenge.

The dataset is provided in two parts Train-1 and Train-2. Train-1 contains 8.6 hours of manually transcribed data from 2017 recordings and Train-2 contains 40.3 hours of data from 2016 and 2018 recordings. For development (dev) and evaluation (eval) sets, the organizers provide around two hours of data each from 2017 recordings. Data statistics, like the number of speakers and words, are presented in Table \ref{tab:corpus}.

As the speech is from different age-groups, the dataset has variations in the speaker's pitch and speaking rates. These variations are reported in Table \ref{tab:speaking_rate}. Comparing pitch, we notice that A1 and A2 groups are similar to some extent but different from the B1 group. While comparing the speaking rate, we observe that A2 and B1 are similar and faster than A1. These differences form the basis for our experiments, where we leverage the pitch and speaking rate to augment speech data.
\renewcommand{\arraystretch}{1.6}
\begin{table}[!t]
\centering
\caption{\label{tab:scoring_eg}The table shows a portion of an utterance from the reference text (Reference) in the evaluation set. The evaluation script removes the words in red in this text to produce the modified text for WER calculations (Modified). We also present the example utterance's hypothesis predicted by our best system before confidence-score-based filtering (Prediction) and the filtered version  (Filtered) produced by dropping the words with very low-confidence scores (marked in blue). The filtering process is described in section \ref{sec:conf_filter}.}
\vspace{-0.3cm}
\begin{tabular}{|l|l|}
\hline
Reference & {\color{red} \textless unk\textgreater} in {\color{red} \textless unk\textgreater} the people i watch the\\
Modified & in the people i watch the\\
Predicted & {\color{blue} @uh} interesting {\color{blue} ping} in {\color{blue} work} people i watch the\\
Filtered & interesting in people i watch the\\
\hline
\end{tabular}
\vspace{-0.6cm}
\end{table}
\subsection{Evaluation procedure}
\label{ssec:scoring}
An interesting aspect of this task is the modified Word Error Rate (WER) metric used to compare ASR performance. Before computing the regular WER, the modified procedure filters out non-English tokens like unknown tokens (e.g. \texttt{<unk>}, \texttt{<unk-it>}, \texttt{<unk-de>} ), disfluencies like false starts (e.g. \texttt{pro-} from \texttt{pro- program}) and filler tokens (e.g. \texttt{@m}, \texttt{@e}) from both the hypotheses and references. Intuitively, removing such words leads to a comparison of only English words to calculate WER. Table \ref{tab:scoring_eg} shows an example of this process.
\section{Baseline systems}
\label{sec:baseline}
The challenge organizers provide a Kaldi toolkit-based recipe to train on the Train-1 portion (9 hours) of their data \cite{Povey_ASRU2011_kaldi}. This setup utilizes MFCC features as input to train TDNN-based acoustic models \cite{PoveyCWLXYK18} on LDA-MLLT+SAT based GMM alignment labels. The recipe also performs speaker adaptation of the acoustic model using i-vectors \cite{SaonSNP13}. The decoding is then performed using a 4-gram maximum entropy language model built using SRILM toolkit \cite{Stolcke02}. The recipe can be further modified to train acoustic and language models on combined Train-1 and Train-2 portions of the data. Training on the combined data leads to large performance improvements in terms of Word-Error-Rate (WER), as noted in Table \ref{tab:dev_am_wer}. We also include two spelling corrections in these results, where the American English spelling \textit{favorite} is replaced by its British variant \textit{favourite} and the word \textit{coca-cola} is split in two words \textit{coca cola}. These spelling corrections help normalize the spelling used across training and development portions of the dataset.
\renewcommand{\arraystretch}{1}
\begin{table}[!t]
\begin{center}
\caption{\label{tab:speaking_rate} The table reports the average pitch and speaking rate among different age-groups: A1 (9-10 years), A2 (12-13 years) and B1 (14-16 years). We measure the speaking rate in words per second for the same duration across the different age groups. The table also shows the number of utterances with false starts in the development set (dev).}
 \vspace{-0.3cm}
 \scalebox{1.00}{
\begin{tabular}{|l|c|c|c|}\hline
Data 			&A1	&A2 & B1 	\\ 
\hline
\rowcolor{lightgray} \multicolumn{4}{|c|}{Pitch scale}\\
\hline
Pitch & 218 & 212 & 194 \\
\hline 
\rowcolor{lightgray} \multicolumn{4}{|c|}{Speaking rate}\\
\hline
Word/sec & 0.614 & 1.159 & 1.036 	 \\ 
No. of words		    & 11501	& 21699 	& 19401			\\ \hline
Duration in sec 		& \multicolumn{3}{c|}{18720} 				\\ \hline
\hline
\rowcolor{lightgray} \multicolumn{4}{|c|}{Partial words}\\
\hline
Utterances \% (dev) & 8.4 & 14.5 & 27.7 \\
\hline
\end{tabular}}
\end{center}
\vspace{-0.8cm}
\end{table}

We also train bidirectional LSTM-based TDNN (TDNN-BLSTM) acoustic models as a comparative baseline system. This model performs worse than the regular TDNN on WER, as shown in Table \ref{tab:dev_am_wer}. However, this system shows benefits when combining with other different acoustic models (Section \ref{sec:sys_comb}) employed in this paper. We also plan to investigate other acoustic model architectures like CNN-TDNN as well for this task. In the rest of the paper, we use the baseline recipe unless specified otherwise.

To handle the speech of different age groups, we also perform Vocal Tract Length Normalization (VTLN) \cite{EideG96} per age-group, which can aid speaker adaptation at the input feature level. Training the TDNN acoustic model with the modified data outperforms the simplistic TDNN baseline (Table \ref{tab:dev_am_wer}).

\section{Data augmentation}
\label{sec:data_augment}
In this paper, we implement a prosody-based data augmentation technique, which we describe in Section \ref{ssec:prosody}. We also contrast this technique to SpecAugment-based data augmentation \cite{ParkCZCZCL19}, which has recently shown benefits for ASR in general.
\subsection{Prosody modification based data augmentation}
\label{ssec:prosody}
We change the pitch scale and the speaking rate systematically to leverage prosodic variation in the children's speech (Section \ref{ssec:data}). This process introduces more acoustic variability to the original children's speech corpora. We then augment the modified data to the original corpora for further system development. Figure \ref{fig:Block_Augmentation} summarizes the augmentation process. Intuitively, increasing acoustic variability adds noise to the input features regularizing the learning process.

To modify pitch and speaking rate, we have explored Time Scale Modification (TSM) based on Real-Time Iterative Spectrogram Inversion with Look-Ahead (RTISI-LA) algorithm \cite{Ahmad2017,kathania_2018,kathania_2018_cssp,4244543}. This algorithm constructs a high-quality time-domain signal from its short-time magnitude spectrum. 

RTISI-LA algorithm scales down the pitch per frame of spectrogram with a factor $s$ ($0 < s < 1$) and upsamples this frame to maintain the original size. Next, the algorithm computes a short-time Fourier transform magnitude (STFTM) of the obtained frame. The STFTM describes the audio signal, perceived in terms of its frequency components, by combining the imaginary and real parts into a single number. The RTISI-LA reconstructs the audio signal from its STFTM through an iterative process. Similarly, the RTISI-LA  \cite{kathania_2018,kathania_2018_cssp,4244543} can vary the speaking-rate by changing the length of the speech signal per unit time by varying the speaking rate factor $\alpha$.

 
Table \ref{tab:dev_am_wer} reports the ASR performance for different prosody-based augmentations made to the original data. These include:
\begin{enumerate}
    \item when only speaking rate (SR) modified data with $\alpha=1.1$ is augmented to the original,
    \item when only pitch scale (P) modified data with $s=0.9$ is augmented to the original,
    \item both SR and P modified data are augmented (SR-P) to the original, and
    \item the original data further augmented with speaking-rate modification with $\alpha=1.2$ and pitch scale modification with $s=0.85$ and added to SR-P (SR2-P2).
\end{enumerate} 
\renewcommand{\arraystretch}{1.2}
\begin{table}
\centering
 \footnotesize
\caption{The table reports WER for various ASR systems. The systems vary in acoustic models, amount of data used for training and acoustic modification applied (if any). For details on augmentation techniques refer to section~\ref{sec:data_augment}. Asterisks (*) denote statistical significance while comparing against TDNN (23.06) using the matched pairs test with $p < 0.001$.   \label{tab:dev_am_wer}}
\vspace{-0.3cm}
\scalebox{1.00}{
\begin{tabular}{ |c|c|c|c| } 
 \hline
 AM & Data size & Acoustic mods. & WER\\ 
\hline \hline 
\rowcolor{lightgray}
\multicolumn{4}{|c|}{Baselines}\\
\hline
 TDNN (9 hrs) & 0.2x & - & 37.75\rlap{$^*$}\\
 TDNN (baseline) & 1x & - &  23.06\\
 TDNN-BLSTM & 1x & - &  29.07\rlap{$^*$}\\
 TDNN & 1x & VTLN & \textbf{22.68} \\
\hline
\rowcolor{lightgray}
\multicolumn{4}{|c|}{SpecAugment}\\
\hline
 TDNN & 2x & SpecAug & \textbf{22.21}\rlap{$^*$}\\
 TDNN & 4x & SpecAug & 22.85 \\
\hline
\rowcolor{lightgray}
\multicolumn{4}{|c|}{Prosodic Modifications}\\
\hline
 TDNN & 2x & Speaking rate (SR) & 22.58 \\
 TDNN & 2x & Pitch (P) & 21.92\rlap{$^*$}\\
 TDNN & 3x & SR-P  & 21.75\rlap{$^*$}\\
 TDNN & 5x & SR2-P2& \textbf{21.58}\rlap{$^*$}\\
\hline
\end{tabular}}
\vspace{-0.5cm}
\end{table}
In cases 1) and 2), augmentation doubles the amount of data. Here the pitch-scale based modifications result in a better performance between the two. We observe subsequent improvements when increasing the data via pooling of data from pitch-scale and speaking rate modifications, i.e., SR-P and SR2-P2. In all these cases, the increased prosodic variability helps to improve ASR performance.
\subsection{SpecAugment}
\begin{figure}[!t]
\centering
 \includegraphics[height=3.5cm, trim={0 0.3cm 0 0}]{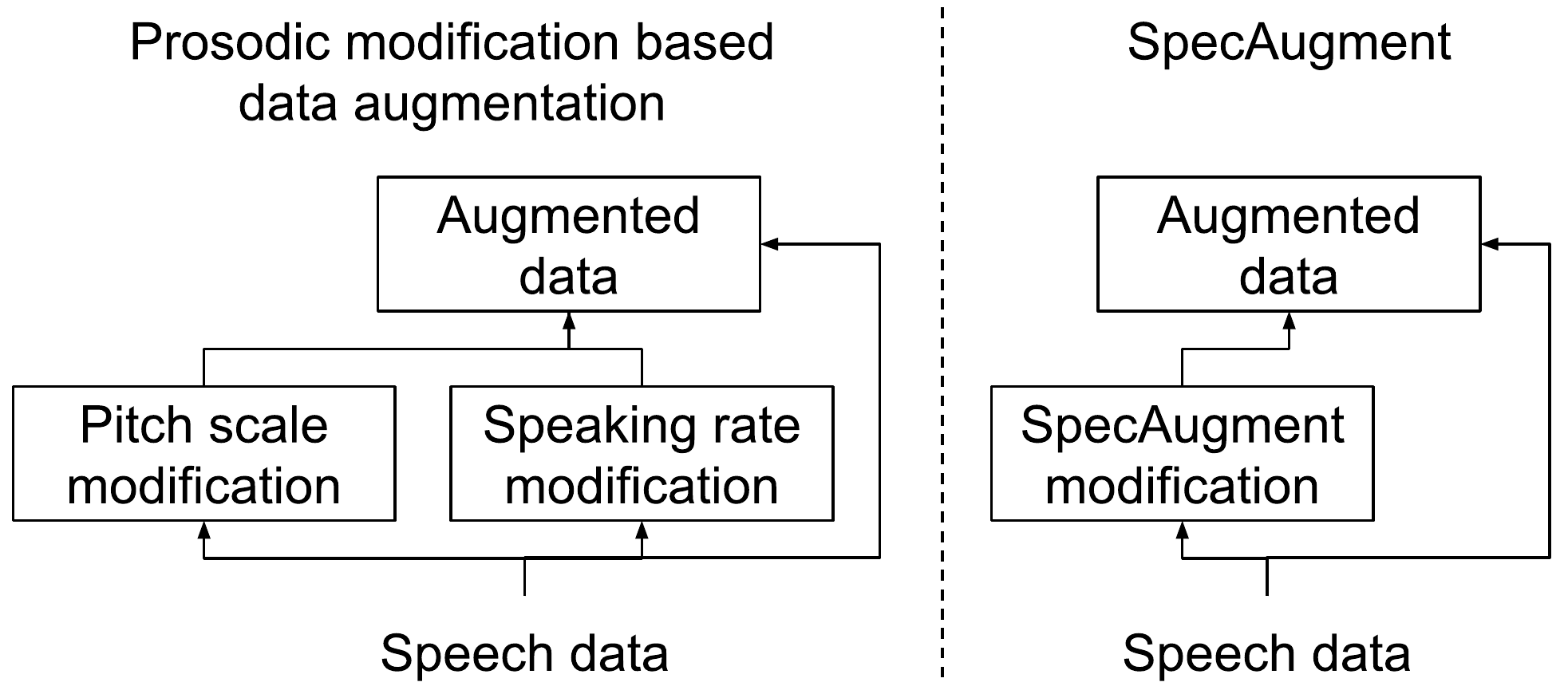}
\caption{The figure displays a block diagram depicting the different data augmentation methods used in our work.}\vspace{0mm}
\label{fig:Block_Augmentation}
\vspace{-0.3cm}
\end{figure}

SpecAugment \cite{ParkCZCZCL19} modifies the input spectrogram by removing time and frequency information randomly. It further warps information across the time axis producing variable speaking rates in different segments of the audio. In our experiments, we use \textit{Librispeech double} augmentation policy, which had performed well on the Librispeech dataset \cite{ParkCZCZCL19}, and applied it directly to the MFCC features to create additional data. In the future, we explore SpecAugment applied to filter bank features as done in the original recipe \cite{ParkCZCZCL19}. Similar to prosody-augmentation, the modified data is augmented to the original for further use in ASR development, as shown in Figure \ref{fig:Block_Augmentation}.

Table \ref{tab:dev_am_wer} shows that doubling the data through SpecAugment (2x SpecAug) improves performance while subsequent increase (4x SpecAug) leads to worse results. This effect is in contrast with Prosody-based augmentation, which shows a consistent improvement with subsequent increase in augmented data size.
\renewcommand{\arraystretch}{1.1}
\begin{table}
\centering
\footnotesize
\caption{This table shows WERs for the TDNN-based system with language models trained using different language modeling text. Asterisks (*) denote statistical significance while comparing against (1) using the matched pairs test with $p < 0.001$. \label{tab:lms}}
\vspace{-0.6cm}
\begin{tabular}{ |c|c|c|c| } 
 \hline
 \# & LM Text & WER & +Conf. Filter\\
 \hline \hline 
 1 &\multicolumn{1}{l|}{Original text} & 23.06 & 22.38\\
 2 & \multicolumn{1}{l|}{\quad+False start noise} & 23.32\rlap{$^*$} & 23.89\rlap{$^*$} \\
 \hline
 3 & \multicolumn{1}{l|}{Unnormalized text} & 23.09 & 22.58 \\
 4 & \multicolumn{1}{l|}{\quad +False start noise} & 23.93 & 23.61\rlap{$^*$}\\
 \hline
 \rowcolor{lightgray}
 \multicolumn{4}{|c|}{Linear combinations}\\
 \hline
 \multicolumn{2}{|c|}{1+2} & 23.00 & 23.61\rlap{$^*$}\\
 \multicolumn{2}{|c|}{1+3} & 22.92\rlap{$^*$} & \textbf{22.32}\\
 \multicolumn{2}{|c|}{1+4} &\textbf{22.83} & 22.51 \\
 \hline
\end{tabular}
\vspace{-0.6cm}
\end{table}

\section{Language modeling}
The organizers normalize 2016 and 2018 transcripts for language modeling (Original) by removing some of the filler tokens. Training language models with the normalized text bias them, as contexts with filler tokens become unseen. The 2017 transcripts are made available in an \textit{unnormalized} form with the filler tokens intact. Along with training language models with normalized text, we also create language models with unnormalized text. Like the baseline recipe, we create a 4-gram maximum entropy model for both normalized and unnormalized text. Table \ref{tab:lms} shows the performance of these two models. We also linearly interpolate these models trained on different corpora, which improves over the constituent models.
\vspace{-0.1cm} 
\subsection{Handling false starts in text}
\label{ssec:partial_word}
In TLT-school corpus, children across different age-groups frequently hesitate while speaking. These false starts are marked as partial words (like \texttt{pro-} in \texttt{pro- program}). Table \ref{tab:speaking_rate} shows the percentage of utterances per age-group containing partial words. We observe that the partial words affect around 10\% portion of the development set. The B1 speakers report the most number of false starts as they usually speak longer sentences having a higher probability of hesitating per sentence.

To handle partial words, we artificially add such words in the language modeling corpus. In this process, we first sample words with at least three characters from the language modeling corpora. The sampled word is split in a random position. Finally, the sampled word is replaced by the partial word and itself (e.g. \texttt{program} $\rightarrow$ \texttt{prog- program}). Then the noised text can be used to build language models. The noised-text models do not perform well compared to the source text models, as shown in Table \ref{tab:lms}. However, they show improvement when augmented with original-text models via linear interpolation. We note that interpolating with original-text models is crucial; otherwise, the noise-text models do not perform well.

\section{Filtering the decoding output}
\label{sec:conf_filter}
In this task, the modified WER (Section \ref{ssec:scoring}) removes non-English words before calculating WER. This removal is to facilitate the comparison of only English words. However, in practice, the ASR system can incorrectly predict non-English tokens as English words and contribute to the error.

In our post-evaluation experiments, we built a filter to remove these words from the decoding output --- the filter inputs the word confidence scores, a combination of acoustic and language model scores. The filter only outputs words that have a confidence score above a certain threshold. We show an example of this filter in action on a segment of text from the evaluation set in Table \ref{tab:scoring_eg}. From the predicted statement (Predicted), the filter can remove incorrectly recognized words in the Filtered statement.  Also, filtering out low-confidence score words helps improve the ASR performance, as shown in Table \ref{tab:lms}. We chose the filter thresholds in the range of $[0,1]$ that produced the best WER on the development set.
\renewcommand{\arraystretch}{1.3}
\begin{table}
\centering
\caption{The table presents WERs of different ASR system combinations on development (dev) and evaluation (eval) sets. $\dagger$ marks our best-submitted system. Post-evaluation, we improved on these results and our best results are marked in boldface. Asterisks (*) denote statistical significant results compared to the baseline using the matched pairs test with $p < 0.001$. \label{tab:sys_comb}}
\vspace{-0.3cm}
\footnotesize
\begin{tabular}{ |c|c|c|p{0.7cm}| } 
\hline
 AM & LM & dev & eval\\
 \hline \hline 
 \rowcolor{lightgray}
 \multicolumn{4}{|c|}{Individual systems}\\
 \hline
 TDNN (baseline) & Original & 23.06  & 20.26 \\
 SR2-P2 & Org.+Unnorm. & 21.45\rlap{$^*$} & 19.71 \\
 SR2-P2 + Conf. Filter & Org.+Unnorm. & 21.20\rlap{$^*$} & 19.10\rlap{$^*$}\\
 \hline
 \rowcolor{lightgray}
 \multicolumn{4}{|c|}{System combinations}\\
 \hline
 TDNN+BLSTM+VTLN & Original & 21.92\rlap{$^*$} & 19.58 \\
 3x(2x SpecAug) & Original & 21.18\rlap{$^*$} & 19.84\\
 SR-P+SR2-P2 & Original & 20.92\rlap{$^*$} & 19.03 \\
 \hline
 \rowcolor{lightgray}
 \multicolumn{4}{|c|}{System combinations with TDNN+BLSTM+VTLN}\\
 \hline
 \multicolumn{1}{|l|}{+3x(2x SpecAug)} & Original & 21.01\rlap{$^*$} & 19.13\rlap{$^*$}\\
 \hline
 \multicolumn{1}{|l|}{\quad+SR-P} &\multicolumn{1}{l|}{Original} & 20.86\rlap{$^*$} & 18.80\rlap{$^*$}\\
 \multicolumn{1}{|l|}{\quad+SR-P} &\multicolumn{1}{l|}{\quad+Unnormalized} & 20.54\rlap{$^*$} & 19.01\rlap{$^*$}\\
 \multicolumn{1}{|l|}{\quad+SR-P} &\multicolumn{1}{l|}{\quad\quad+Noise} & 20.43\rlap{$^*$} & 18.71\rlap{$^{*\dagger}$} \\
 \hline
 \multicolumn{1}{|l|}{\quad\quad+SR2-P2} & \multicolumn{1}{l|}{Original} & 20.60\rlap{$^*$} & 18.68\rlap{$^*$} \\
 \multicolumn{1}{|l|}{\quad\quad+SR2-P2} &\multicolumn{1}{l|}{\quad+Unnormalized} & 20.31\rlap{$^*$} & 18.58\rlap{$^*$} \\
 \multicolumn{1}{|l|}{\quad\quad+SR2-P2} &\multicolumn{1}{l|}{\quad\quad+Noise}& 20.09\rlap{$^*$} & 18.42\rlap{$^*$} \\
 \hline
 \multicolumn{1}{|l|}{\quad\quad\quad+Conf. Filter} & Org.+Unnorm.+Noise & 19.86\rlap{$^*$} & \textbf{17.99}\rlap{$^*$} \\
 \hline
 \end{tabular}
 \vspace{-0.6cm}
\end{table}

\section{System combinations}
\label{sec:sys_comb}
As no external resources are used to create our ASR systems, we submit our systems as part of the \textit{closed track} for the shared task on recognizing non-native children's speech. On this task, we report the best individual and combined systems' WER as chosen on the development (dev) set and evaluated on the test set (eval) in Table \ref{tab:sys_comb}. For evaluation, we retrained all our systems on the training and development set. Among the individual systems, the ASR trained on the most amount of augmented data (SR2-P2) achieves the best result. This system also applies the word confidence score filtering and language models built using original 2016-2018 (Original) and unnormalized 2017 (Unnormalized) transcripts.

Earlier, increasing the amount of SpecAugment-based data (4x SpecAug) did not result in an improvement. Nevertheless, combining three different 2x SpecAug ASR systems, labeled as 3x(2x SpecAug), leads to improvement over the individual 2x SpecAug system. Combining both the prosody-augmentation based systems (SR-P+SR2-P2) leads to the best data augmentation based system.

During the evaluation period, we submitted a system combination of 3x(2x SpecAug) with SR-P. This ASR system also utilizes an interpolated language model where the constituents are trained on the original normalized text, unnormalized text and noised unnormalized (noise) text. For individual systems, the noised-text-based language models did not improve performance compared to using just original and unnormalized text-based models but turned out to be essential for building combined systems. In our post evaluation, we improved this system by combining it with SR2-P2, which adds more prosody-augmentation and applies word score confidence filtering. These additions helped to achieve our best WER of 17.99. 
\section{Related work}
In the context of children speech, prosodic features and modifications are well studied \cite{8049381,Ahmad2017,kathania_2018_cssp,kathania_2018_ICSSP,SHAHNAWAZUDDIN2020213}. Prior work \cite{SHAHNAWAZUDDIN2020213} has leveraged similar prosody modifications for data augmentation in children ASR achieving substantial gains in performance. These benefits also inspire our solution. Though, unlike prior work \cite{SHAHNAWAZUDDIN2020213}, which used a glottal-closure-instants-based modification, we use a simpler TSM based algorithm to modify the prosodic parameters like pitch and speaking rate. 

In the context of language modeling, quite a few researchers \cite{PrylipkoVSW12,7804538,StolckeS96,SiuO96,LiuSS03,MonizTM09} have studied disfluencies like false starts, filled pauses and repetition in textual data. Some of these work \cite{PrylipkoVSW12,7804538} have noted that modeling disfluencies can be beneficial for language modeling. In the same vein, our work focuses on modeling false starts in text. Most similar to our work, \cite{7804538} introduces disfluencies to clean text and uses the processed text for the language model. They, however, use an existing set of disfluencies to be introduced in the text. In contrast, we split words in the text to introduce new disfluencies.
\section{Concluding remarks}
\label{sec:conclusion}
In this work, we presented AaltoASR's system for the task of recognizing non-native children's speech. We focused on applying a data augmentation-based approach. We leveraged prosody- and SpecAugment-based data augmentation to augment the limited training data for building acoustic models. Compared to SpecAugment, prosody-based augmentation achieved better results. Additionally, prosody-based augmentation showed improvement when increasing the amount of augmentation data, whereas increasing the amount of augmented data for SpecAugment led to worse results. 

We also modeled false starts (partial words) in the text to augment the language modeling corpora for training. Adding the partial-word noise improved the ASR performance of linearly interpolated models compared to the vanilla language models. In our post-evaluation experiments, we developed a filtering mechanism to remove low confidence scores from the decoded output, which helped improve the ASR performance. Finally, we performed a system combination of the techniques developed in this work to achieve the best result.

\vspace{-0.2cm} 
\section{Acknowledgements}
This work was supported by the Academy of Finland (grant 329267) and the Kone Foundation.  The computational resources were provided by Aalto ScienceIT.

\ninept
\bibliographystyle{IEEEtran}
\bibliography{references_ev}

\end{document}